%% file: main.tex
\documentclass[10pt,twocolumn,letterpaper]{article}

\usepackage[pagenumbers]{cvpr}   %

\usepackage{graphicx, subcaption}
\usepackage{booktabs, multirow, makecell}
\usepackage[export]{adjustbox}
\usepackage{pgfplots, pgfplotstable}
\pgfplotsset{compat=1.18}
\usepackage{xcolor}
\usepackage{listings}
\usepackage{cuted}

\RequirePackage{xspace}
\makeatletter
\DeclareRobustCommand\onedot{\futurelet\@let@token\@onedot}
\def\@onedot{\ifx\@let@token.\else.\null\fi\xspace}
\def\eg{\emph{e.g}\onedot}

\makeatother

\newif\ifshowchanges
\showchangesfalse  %
\newcommand{\changes}[1]{%
  \ifshowchanges\textcolor{blue}{#1}\else#1\fi
}

\definecolor{cvprblue}{rgb}{0.21,0.49,0.74}
\usepackage[pagebackref,breaklinks,colorlinks,allcolors=cvprblue]{hyperref}

\def\paperID{arXiv}
\def\confName{CVPR}
\def\confYear{2026}

\begin{document}

\title{MultiCube: \\ Compositional 3D Generation With Part-Level Semantic and Spatial Control}

\author{
Ava Pun\textsuperscript{1,2} \qquad Kangle Deng\textsuperscript{1} \qquad Yiheng Zhu\textsuperscript{1}
\\
Jun-Yan Zhu\textsuperscript{2} \qquad Maneesh Agrawala\textsuperscript{1,3} \qquad Tinghui Zhou\textsuperscript{1}
\\
\textsuperscript{1}Roblox \qquad \textsuperscript{2}Carnegie Mellon University \qquad \textsuperscript{3}Stanford University
}

\maketitle

\input{figTex/teaser}
\input{sections/0_abstract}

\input{sections/1_intro}
\input{sections/2_related_work}
\input{sections/3_method}
\input{sections/4_experiments}
\input{sections/9_conclusion}

\clearpage

\bibliographystyle{ieeenat_fullname}
\bibliography{main}

\clearpage

\appendix
\input{sections/A_supplemental}

\end{document}

%% file: figTex/teaser.tex
\begin{strip}\centering
\vspace{-30pt}
    \includegraphics[width=\linewidth]{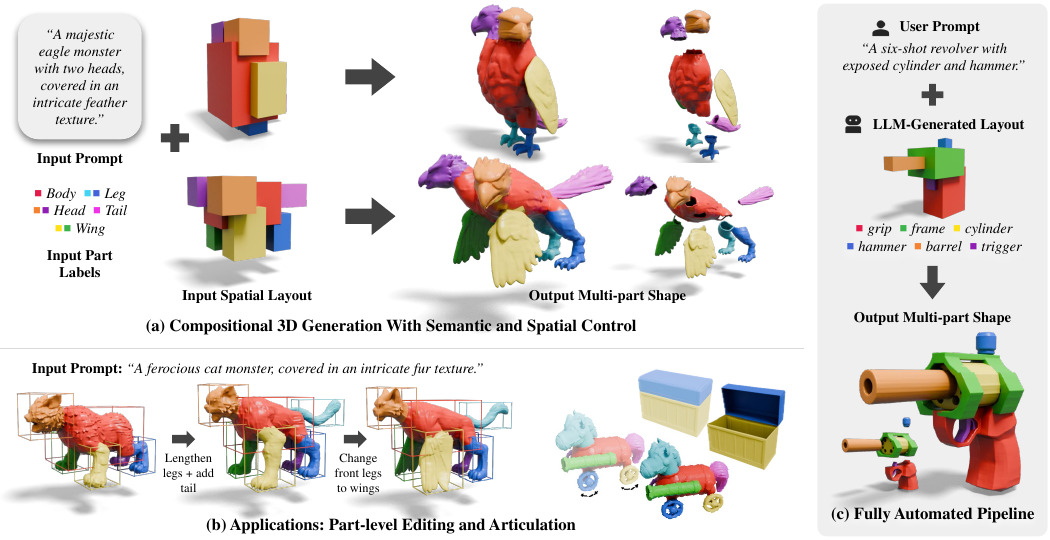}
    \captionof{figure}{\textbf{We introduce MultiCube, a framework for \textit{controllable} compositional 3D generation.} Conditioned on an input text prompt, $n$ part labels, and $n$ part bounding boxes, MultiCube outputs an $n$-part 3D shape aligned with the given semantic and spatial constraints (a). The generated shapes allow for part-level editing and adding articulations (b). By integrating with LLM-based layout creation, the generation pipeline can be made fully automatic (c).
    Please check our project page: \url{http://multi-cube.github.io}.
    }
    \label{fig:teaser}
\end{strip}

%% file: sections/0_abstract.tex
\begin{abstract}
    Digital 3D objects used in games and animation are often required to be \textit{compositional}; that is, decomposed into semantically meaningful parts. Recent 3D generation methods can produce high-quality compositional objects conditioned on image or text prompts. Yet, such global conditioning lacks the precise part-level controllability required for professional creative workflows. To address this, we introduce \textit{MultiCube}, a novel compositional 3D generation method that provides explicit, independent control over both the semantics and spatial arrangement of each part. MultiCube takes as input a global text prompt, a text schema specifying the desired parts, and a spatial layout indicating the bounding boxes of the parts in the given schema. It outputs a 3D object composed of distinct meshes, one per specified part, that adhere to the given semantic and spatial conditions. Our approach employs a two-stage diffusion process, first generating a schema- and layout-aligned monolithic mesh, then decomposing the mesh into individual parts simultaneously. A novel \textit{Part Layout Adapter} is used to encode per-part conditions \changes{independently of the other parts.} Experiments demonstrate that our method can generate high-quality compositional 3D objects with precise part-level control, including those with unique layouts difficult to achieve with text or image prompting alone.

\end{abstract}

%% file: sections/1_intro.tex
\section{Introduction}

Digital 3D assets are crucial for fields such as animation, gaming, robotics, and virtual reality. The creation of 3D content has traditionally been a time-consuming process requiring much manual work and experience. Fortunately, recent advances in generative AI enable even non-expert users to create high-quality 3D assets from simple text or image prompts, democratizing the creation process and accelerating creative pipelines. Although many prior works focus on generating \textit{monolithic} objects, in which geometry is presented holistically as a single mesh~\cite{hunyuan3d,hi3dgen,triposg,trellis}, some recent methods have tackled the task of generating \textit{compositional} objects, those composed of several semantically meaningful parts~\cite{partcrafter, partpacker, unipart, fullpart, omnipart, moca, partgen, part123}. Compositional 3D generation more closely mirrors the traditional creation process, and it facilitates many downstream capabilities, including part-level editing, physical simulation, and animation.

\input{figTex/semantic-control}
\input{figTex/baseline-failure}

However, current compositional 3D generation methods still lack the controllability required for professional workflows. In modern games or animation, character and object behaviors often operate on a fixed set of parts. For example, a developer's game logic might expect a character model to consist of a body part, two arm parts, and two leg parts. To be functional, a compositional 3D object must adhere to a ``part schema'' indicating the number and semantics of its constituent parts. Yet 3D generation methods offer limited control for aligning the generated parts with such a schema. Some methods~\cite{partpacker} produce an arbitrary set of parts; the user cannot decide how many parts to generate or what those parts should be. Other methods~\cite{partcrafter, moca, unipart} allow the user to specify only the number of parts. Some methods allow the user to specify parts by segmenting an input 2D image~\cite{omnipart, partgen, part123} or 3D shape~\cite{autopartgen, holopart, x-part}, but this process can be tedious and requires the input image or shape to contain all the necessary parts. With a 2D input, it is also impossible to control the parts occluded from the input view. This lack of detailed part-level control limits the utility of compositional 3D generative models in practice.

Furthermore, most text- or image-conditioned 3D generation methods offer limited artistic control, as 3D objects cannot be completely described by these modalities alone. In particular, detailed shapes and spatial arrangements are cumbersome to express via natural language and cannot be fully captured by a single-view image or 2D layout. Text-to-image models struggle to follow spatially complex prompts; additionally, while there has been much research into generating images guided by 2D bounding box layouts~\cite{gligen,migc,creatilayout}, these methods are optimized for multi-object scenes and fail to produce single objects with specific part layouts (\cref{fig:baseline-failure}). To give users better control over the shape of generated 3D objects, previous works have explored various 3D-aware conditions, including single bounding boxes~\cite{clay}, voxels \cite{art-deco, clay}, and primitives~\cite{coin3d}. However, these conditions are not \textit{part-aware}; though users can change the overall shape of the object, they have limited influence over the shape and position of individual parts.

We argue that controllable compositional 3D generation should have two capabilities. First, it should allow users to explicitly specify \textit{part semantics}. As different downstream tasks require diverse part decompositions, this control must be flexible and open-vocabulary. Second, the generation should allow for \textit{part-aware layout control}. Users should be able to specify the location and dimensions of each part independently of the remaining parts, granting them fine-grained control over the shape of the output. 

To support these capabilities and help users better realize their creative visions, we present \textit{MultiCube}, a novel framework for compositional 3D generation. MultiCube takes as input a global text prompt describing the overall object (\eg \textit{``a mountain bike''}), a text schema specifying the desired parts (\eg \textit{``body'', ``front wheel'', ``back wheel''}), and a spatial layout indicating the bounding boxes of the parts in the given schema. It produces several distinct part meshes, one per specified part, that fit together to make a coherent object aligned with the input text prompt and multi-box layout (\cref{fig:teaser} (a)). Importantly, the spatial characteristics of each labeled part are \textit{explicitly controlled via its associated bounding box}, rather than being inferred from the overall spatial layout; users can specify different part semantics for the same spatial layout, and vice versa (\cref{fig:semantic-control}). This level of control enables users to create 3D objects with distinct layouts, which is difficult to achieve with text prompting alone. Part layouts can be specified either manually, or using a large language model (LLM) for a fully automated generation pipeline (\cref{fig:teaser} (c)).

MultiCube employs a two-stage diffusion-based 3D generation process. In Stage 1, it is conditioned on the global prompt, part schema, and multi-box layout to produce a monolithic mesh that contains all the required parts in the specified positions. In Stage 2, it decomposes the mesh into the specified parts, generating the parts simultaneously with cross-part attention to ensure global consistency. MultiCube utilizes a novel \textit{Part Layout Adapter} during Stage 1 to encode the semantic and spatial conditions for each part. This module processes each part separately, preventing geometric constraints from entangling and providing a strong conditioning signal for monolithic shape generation.

In summary, our contributions are as follows.
\begin{itemize}
    \item We propose a new problem formulation for part-layout-conditioned compositional 3D generation, encompassing both semantic \textit{and} spatial part-level control.
    \item We introduce MultiCube, a novel method for generating compositional 3D objects conditioned on a text prompt, part labels, and per-part spatial conditions. MultiCube utilizes a Part Layout Adapter to encode part conditions \changes{independently, providing a per-part conditioning signal for monolithic shape generation.}
    \item We demonstrate that our method can generate high-quality compositional 3D objects with precise part-level control.
\end{itemize}

%% file: figTex/semantic-control.tex
\begin{figure}[tb]
    \centering
    \includegraphics[width=\linewidth]{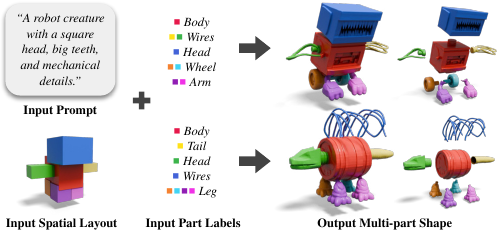}
    \caption{\textbf{Part-level semantic control.} Users can specify \textit{different part semantics} for the same spatial layout, resulting in unique output designs.}
    \label{fig:semantic-control}
\end{figure}

%% file: figTex/baseline-failure.tex
\begin{figure}[tb]
    \centering
    \includegraphics[width=\linewidth]{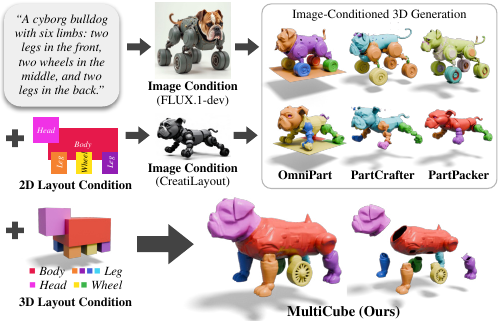}
    \caption{Text- or 2D layout-conditioned image generation models, and the image-conditioned 3D generation methods that rely on their output, struggle to follow prompts with complex part layouts. With 3D layout conditioning, MultiCube can generate shapes aligned with precise part layouts.}
    \label{fig:baseline-failure}
\end{figure}

%% file: sections/2_related_work.tex
\section{Related Work}

\paragraph{3D Shape Generation.}

Early progress in 3D generation was driven by 2D-to-3D lifting. Methods like DreamFusion~\cite{dreamfusion} introduced Score Distillation Sampling (SDS) to optimize implicit representations (\eg, NeRFs~\cite{nerf}) using 2D diffusion priors. Many follow-up works~\cite{magic3d,prolificdreamer,zero123,fantasia3D} adopted this to compensate for limited 3D data. However, relying on view-dependent image supervision provides only partial structural constraints and lacks explicit mechanisms for part control.

With the advent of large-scale 3D datasets like Objaverse~\cite{objaverse} and Objaverse-XL~\cite{objaversexl}, 3D-native generation has emerged as a powerful alternative. One approach utilizes compact latent spaces, beginning with the VecSet representation~\cite{3dshape2vecset} and scaling up in recent works~\cite{michelangelo,craftsman3d,cube3d,step1x3d,lattice,clay,triposg,hunyuan3d,oat3d} for end-to-end 3D shape generation.
A parallel line of work leverages sparse voxel grids ~\cite{trellis,trellis2,sparseflex,sparc3d,direct3ds2,hi3dgen}
to support localized, high-resolution geometry synthesis. Despite these architectural advances, both VecSet and sparse-voxel methods generally output monolithic meshes and do not support semantic part-level decomposition or structural control.  To address this limitation, our method builds upon the VecSet representation, extending it to enable both semantic and spatial part-level control.

\paragraph{Part-level 3D Generation.}

Early approaches to part-level 3D generation primarily focused on compositional latent representations, such as SPAGHETTI~\cite{spaghetti} and StructureNet~\cite{structurenet}, which learned part-aware latent spaces to enable part-aware editing. However, these methods often struggled to generate highly detailed, diverse geometries. Consequently, many subsequent part-level 3D generation methods transitioned to a segmentation-then-reconstruction pipeline, leveraging external segmentation priors to partition an object before reconstructing each part individually. HoloPart~\cite{holopart} segments a 3D object and completes each component via latent denoising, while PartGen~\cite{partgen} employs multi-view diffusion to generate 2D part segmentation maps for per-part reconstruction. Similarly, AutoPartGen~\cite{autopartgen} autoregressively generates parts one at a time conditioned on 2D part masks. These multi-stage pipelines are often prone to error propagation from inaccurate boundaries. 

Recent works instead focus on direct part-aware generation without explicit segmentation priors. PartCrafter~\cite{partcrafter} jointly denoises multiple parts within a compositional VecSet-based latent space, and PartPacker~\cite{partpacker} uses a dual volume packing strategy to handle arbitrary part counts within fixed output lengths. However, both methods offer limited part-level controllability beyond specifying the total number of parts, restricting their practical applicability.
To explicitly manage part layouts, FullPart~\cite{fullpart} plans layouts via implicit box diffusion before generating parts in separate full-resolution grids, and UniPart~\cite{unipart} generates unified geometry-segmentation latents before performing dual-space part diffusion.
OmniPart~\cite{omnipart} first plans 3D bounding boxes then synthesizes all parts using the TRELLIS~\cite{trellis} architecture. However, its generated geometries often fail to conform to the input masks.
Closely related to our method, CubePart~\cite{cubepart} introduces \textit{semantic} part-level control through a two-stage approach, but lacks the fine-grained \textit{spatial} alignment required for complex layouts. In contrast, our method achieves precise and reliable part-level control, both spatially and semantically.

\paragraph{Spatially-controlled 3D Generation.} 

Providing intuitive spatial control over the generation process is crucial for practical 3D modeling workflows. Several recent 3D-native generative models, including CLAY~\cite{clay}, HunyuanStudio~\cite{hunyuanstudio}, and Cube3D~\cite{cube3d}, have introduced spatial control by conditioning the generation process on global bounding boxes.  While this effectively constrains the overall dimensions and scale of the generated object, the control offers no influence over the internal structure or individual components. To achieve finer spatial guidance, ArtDeco~\cite{art-deco} and Coin3D~\cite{coin3d} allow users to condition the generation process using structural primitives or coarse voxel representations.  Although these methods successfully enforce more detailed spatial layouts, they still treat the generated output as a monolithic mesh without explicit part-level semantic control. Furthermore, BANG~\cite{bang} has explored part-level bounding box conditions but assumes a monolithic mesh as input, which is not guaranteed to contain all desired parts in practice. In contrast to these approaches, which either provide only global constraints, lack semantic decomposition, or rely on mesh input, our method generates part-based meshes from text descriptions while simultaneously offering precise control via spatial bounding boxes and explicit semantic conditions for each individual part.

%% file: sections/3_method.tex
\input{figTex/overall-arch}

\section{Controllable Compositional 3D Generation}

Our goal is to generate a compositional 3D object given (1) a global text prompt $p_\text{global}$, (2) a sequence of text part labels $p_1,\dots,p_n$, and (3) a correlated sequence of axis-aligned bounding boxes $b_1, \dots, b_n \in \mathbb{R}^6$ for each part label, specified by their center in $\mathbb{R}^3$ and dimensions in $\mathbb{R}^3$. To achieve this goal, we present MultiCube, our controllable compositional 3D generation method. MultiCube builds upon the architecture of CubePart~\cite{cubepart}, a VecSet diffusion-based method that includes two stages: (Stage 1) monolithic mesh generation and (Stage 2) multi-part mesh generation. 
Specifically, Stage 1 generates a single, monolithic mesh that aligns with the provided part-level bounding boxes, while Stage 2 decomposes this object into individual mesh components.
To enable part-level spatial control, we introduce a Part Layout Adapter and a lightweight bounding box encoder to inject spatial conditions into Stages 1 and 2, respectively. An overview is shown in \cref{fig:overall-arch}.

\subsection{Preliminaries: VecSet Diffusion and CubePart}

VecSet diffusion models~\cite{craftsman3d,triposg,clay,hunyuan3d} are a family of latent diffusion models that generate 3D shapes encoded as sets of unordered latent vectors (VecSets). They typically begin by learning a latent shape representation with a transformer-based Variational Autoencoder (VAE)~\cite{3dshape2vecset}, then train a diffusion model to generate shape latents from noise.

CubePart~\cite{cubepart} is a VecSet-based method that generates multi-part 3D objects conditioned on a text prompt and sequence of part labels. For VAE encoding, it follows the method introduced in~\cite{3dshape2vecset}, where a point cloud with normals is uniformly sampled from the object surface, then encoded by a Perceiver~\cite{jaegle2021perceiver}-based encoder. A Cross Attention layer is designed to compress the point cloud into a fixed length sequence, where the query is down-sampled from the point cloud or a set of learnable vectors. To reconstruct the original 3D object, the latent sequence is first decoded into a signed distance field, then converted into a triangle mesh via Marching Cubes.

CubePart employs a two-stage generation method, using two Multi-Modal Diffusion Transformers (MM-DiTs)~\cite{stable-diffusion-3} for monolithic and multi-part shape latent generation, respectively. In Stage 1, the first DiT generates a monolithic mesh conditioned on an overall text prompt and semantic part labels, merging both prompt and labels into a single text prompt and encoding the resulting text condition into a latent sequence using Qwen-VL~\cite{qwen-vl}. In Stage 2, the second DiT decomposes the monolithic mesh into parts, each part conditioned on a text part label encoded with Qwen-VL. It generates the part latents simultaneously, using cross-part attention blocks to ensure global consistency. This two-stage pipeline takes advantage of the vast amount of monolithic mesh data compared to multi-part data; Stage 1 learns to generate diverse shapes trained on large amounts of single-mesh data, while Stage 2 can learn the easier task of part decomposition from a smaller amount of part-level data.

\subsection{Injecting Spatial Conditions}

\input{figTex/part-layout-adapter}

While the base CubePart model effectively generates compositional 3D objects, it lacks spatial controllability. MultiCube introduces spatial controls by injecting part-level bounding box conditions into both stages of the generation pipeline.

\paragraph{Stage 1: Monolithic Generation.} During the monolithic shape generation stage, CubePart introduces part-level semantic conditions by encoding the part labels jointly with the overall text prompt. However, this combined encoding makes it difficult for the model to associate parts with specific bounding boxes when introducing part-level spatial conditions. Hence, instead of using a combined text encoding, we introduce a novel Part Layout Adapter (\cref{fig:part-layout-adapter}) that fuses semantic and spatial conditions independently for each part.

For each specified part, the adapter takes its semantic label and corresponding bounding box as input, processing them to produce a localized sequence of part latents. The adapter processes each part independently of the other parts, preventing part information from being inappropriately entangled. Before being provided to the adapter, the part label, specified in the format \textit{``Part \{i\}: \{part name\}''}, is encoded using Qwen-VL~\cite{qwen-vl}. To aid the network in learning fine-grained positional variations, the bounding box goes through an 8-frequency Fourier embedding before being linearly projected into the text latent space and concatenated to the part label embedding. The concatenated result is fed into a 2-block Q-Former~\cite{blip-2}, which produces a sequence of 16 part latents. The resulting latent sequences for all parts are concatenated to the global text embeddings, then fed into the monolithic shape generation DiT.

To be precise, denote the adapter by $A$, the Qwen-VL embedder by $Q$, the Fourier embedding by $F$, and the linear layer by $W_1$. For each part $i$ with label $p_i$ and bounding box $b_i$, the adapter encodes it into a latent sequence $h_i$ as follows:
\begin{align*}
    h_i &= A(Q(p_i) \oplus W_1(F(b_i)))
,\end{align*}
where $\oplus$ denotes the concatenation operator. Then the monolithic shape DiT, denoted by $T_1$, takes the noisy latent $z_{\text{mono},t}$ at timestep $t$ alongside the concatenated text and part conditions to predict the denoised monolithic latent:
\begin{align*}
    z_\text{mono} &= T_1(z_{\text{mono},t}, t, Q(p_\text{global}) \oplus h_1 \oplus h_2 \oplus \dots \oplus h_n)
.\end{align*}
Conditioned on the part latents produced by the adapter, the Stage 1 model learns to produce a monolithic shape that includes all of the given semantic features and adheres to the part layout.

\paragraph{Stage 2: Multi-part Generation.} After generating the monolithic shape, MultiCube decomposes the shape into its constituent parts. While the base CubePart model conditions this decomposition solely on semantic labels, it is beneficial to include spatial priors as well, since several parts may have identical or similar semantic labels (\eg multiple legs) and must be differentiated by their bounding boxes. The rough part geometry is already present in the monolithic mesh following the spatial conditions; it must only be completed during part decomposition. Hence, a heavy adapter is not necessary for this stage. Instead, we use a lightweight embedding: for each part, we encode its corresponding bounding box with an 8-frequency Fourier embedding followed by a learnable linear projection, then concatenate it to the text part label provided as conditioning to the multi-part DiT. Guided by semantic and spatial conditions, the Stage 2 model learns to generate a shape matching the desired part decomposition; cross-part attention ensures that the parts are globally coherent.

To be precise, denote the linear layer by $W_2$ and the multi-part DiT by $T_2$. The condition $c_i$ for part $i$ is encoded as follows:
\begin{align*}
    c_i &= Q(p_i) \oplus W_2(F(b_i))
.\end{align*}
Then, the denoising step for the part latent $z_{i,t}$ corresponding to part $i$ and timestep $t$ depends on the monolithic shape $z_\text{mono}$, part condition $c_i$, and latents of all other parts, denoted as $z_{\sim i,t}$:
\begin{align*}
    z_{i,t-1} &= T_2(z_{i,t}, t, z_\text{mono}, c_i, z_{\sim i,t})
.\end{align*}

\subsection{Training}

We take advantage of CubePart's strong 3D generation capabilities by starting with a pre-trained CubePart model~\cite{cubepart} and fine-tuning it for the spatially-guided 3D generation task. To avoid catastrophic forgetting of the pre-trained shape priors, fine-tuning proceeds in two phases. First, only the adapter is trained while the remaining model weights are frozen; then, all weights are trained jointly.

We adopt the flow matching objective~\cite{flow-matching} for training, and we train the Stage 1 and Stage 2 models separately using the same loss function. During training, a shape latent $z_0$ is sampled from the dataset $\mathcal{D}$ and a noisy latent $z_1$ is sampled from the standard multivariate normal distribution $\mathcal{N}(0,\mathbf{I})$. A timestep $t$ is sampled from a logit-normal distribution shifted with a factor of 4, following prior work~\cite{li2024alleviating}. The model input latent at timestep $t$ is defined as $z_t = tz_0 + (1-t)z_1$. Then, our training loss is defined as follows:
\begin{align*}
    \mathcal{L} &= \mathbb{E}_{(z_0,c) \sim \mathcal{D},z_1,t} \| f_\theta(z_t,t,c)-(z_1-z_0) \|^2
,\end{align*}
where $c$ denotes the conditions associated with the sampled shape, and $f_\theta$ denotes the network with trainable parameters $\theta$.

For our training data, we refine the dataset introduced in~\cite{cubepart}, which consists of 510k 3D assets and 2.96M parts. To mitigate collisions and overlapping parts, 
\changes{we filter out training samples containing parts whose volumetric intersection over union, computed on a $64^3$ occupancy grid, is at least 0.3,} leaving approximately 81\% of the original assets.

%% file: figTex/overall-arch.tex
\begin{figure*}[!tbp]
    \centering
    \includegraphics[width=\linewidth]{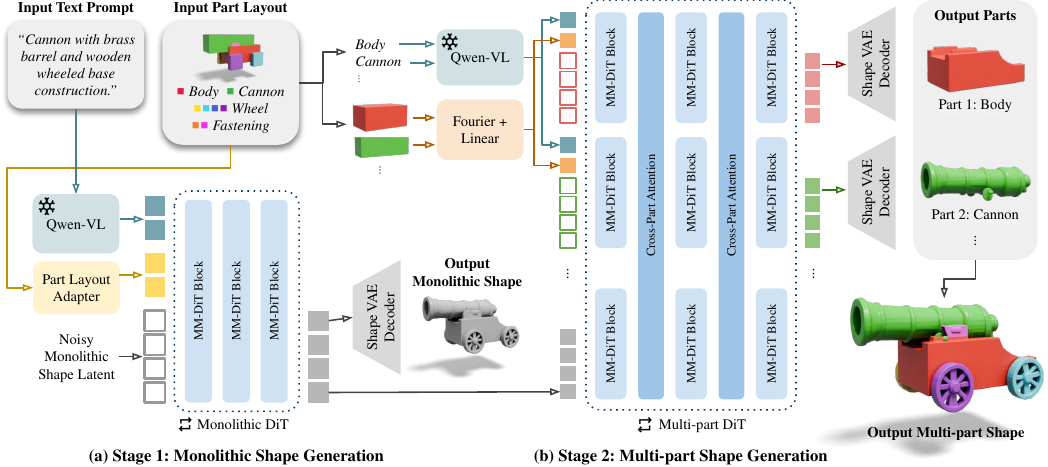}
    \caption{\textbf{MultiCube uses a two-stage diffusion-based architecture.} In Stage 1, it synthesizes a monolithic object conditioned on the input text prompt and part layout (a). In Stage 2, it takes the object latents produced in Stage 1 and decomposes them into multiple shapes, one for each specified part, guided by the part labels and bounding boxes (b). Spatial conditions are encoded with a Part Layout Adapter in Stage 1 and a lightweight embedding in Stage 2.}
    \label{fig:overall-arch}
\end{figure*}

%% file: figTex/part-layout-adapter.tex
\begin{figure}[tb]
    \centering
    \includegraphics[width=\linewidth]{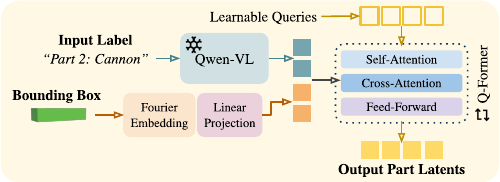}
    \caption{\textbf{Part Layout Adapter.} The adapter encodes each (part label, bounding box) condition pair into a set of part latents. The label is encoded using Qwen-VL and the bounding box goes through a Fourier embedding before being transformed into a latent sequence using a Q-Former~\cite{blip-2}.}
    \label{fig:part-layout-adapter}
\end{figure}

%% file: sections/4_experiments.tex
\input{figTex/full-ablation}

\input{figTex/full-quantitative}

\section{Experiments}
\label{sec:experiments}

We evaluate MultiCube's ability to generate high-quality compositional 3D objects aligned with the given part semantics and spatial conditions.

\subsection{Technical Details}

All training is parallelized across 24 NVIDIA H200s, with the global training batch sizes for Stages 1 and 2 being 768 and 72, respectively. The Part Layout Adapter has 22M trainable parameters, while the Stage 1 and 2 DiTs have 1.9B parameters. We initialize the model with pre-trained CubePart weights and train the adapter for 20 epochs, with all other weights being frozen; then, we fully train both Stage 1 and 2 models for 30 epochs each. We use the AdamW optimizer~\cite{adamw} with no weight decay and $\beta$ values of 0.9 and 0.99. The base learning rate is set to $10^{-4}$, governed by a linear warm-up schedule for the first 2000 iterations. Inference time is \textasciitilde{}1s for Stage 1 and \textasciitilde{}1s per part for Stage 2.

\subsection{Evaluation Dataset}

To evaluate both \changes{Stage 2 and the full pipeline} of our method, we use the PartObjaverse-Tiny dataset~\cite{sampart3d}, containing 200 meshes with human-annotated part labels uniformly sampled from Objaverse~\cite{objaverse}. These meshes are excluded from our training set. We use an instance-level part segmentation and filter out samples with more than eight parts, resulting in 73 evaluation samples.

For full pipeline evaluation, we additionally craft a custom evaluation dataset that includes diverse text prompts, with part labels and bounding box layouts associated with each prompt. First, we create a list of 110 object types (\eg \textit{``fish'', ``motorcycle'', ``lantern''}) and write a list of parts for each type (\eg, a lantern has the parts \textit{``handle ring'', ``housing'', ``glass panes'', ``flame/bulb''}). We then instruct the large language model GPT-5.1 to generate a short, medium, and long text prompt for each object type, \changes{yielding 330 prompts in total, and} devise a suitable part bounding box layout for each prompt. The model is instructed to produce a logically coherent set of bounding boxes within a bounding volume of $[-1.0, 1.0]$. \changes{We apply an automatic connectivity check that rejects layouts containing disconnected bounding boxes; 104 of the 330 layouts are rejected, leaving 226 samples.} Further details are provided in the supplementary.

\input{figTex/stage2-qualitative}

\subsection{\changes{Multi-part Generation (Stage 2) Evaluation}}
\label{sec:stage-2-evaluation}

\input{tables/stage2-quantitative}

We compare our multi-part generation stage against PartCrafter~\cite{partcrafter}, PartPacker~\cite{partpacker}, OmniPart~\cite{omnipart}, FullPart~\cite{fullpart}, HoloPart~\cite{holopart}, and CubePart~\cite{cubepart}. 
\changes{Each baseline is given the strongest conditioning it can accept, so that the comparison is not confounded by missing inputs.}
PartCrafter and PartPacker offer image conditioning only. OmniPart and FullPart accept spatial conditions; we additionally give OmniPart a ground-truth 2D segmentation and 3D voxels. We use PatchAlign3D~\cite{patchalign3d} to segment input meshes based on semantic labels to be given as input to HoloPart, making HoloPart one of two baselines that accepts part-level semantic conditions, alongside CubePart. 
Please refer to the supplementary material for more details about the baseline inputs.
Unlike baselines, our method offers both spatial and semantic control.

To assess shape quality, we compute two metrics: Chamfer Distance (CD) and F-score with a threshold of 0.1. We evaluate both metrics at both the part and holistic (whole mesh) levels. To assess alignment with the input part layout, we compute the IoU between the part and input bounding boxes. All shapes are normalized to a $[-1, 1]^3$ box before comparison. Note that most baselines output an unordered set of parts, which must be matched with reference parts for evaluation. We perform this matching greedily, repeatedly matching the best-scoring pair of reference and generated parts.

As seen in \cref{tab:stage2-quantitative}, our method strongly outperforms all baselines in part-level metrics and remains competitive in holistic-level metrics. \cref{fig:stage2-qualitative} illustrates our method's ability to disentangle complex part layouts and produce high-quality part-level objects.

\input{figTex/full-qualitative}

\subsection{Full Pipeline Evaluation}

\changes{We first ablate our design choices, then compare the full pipeline against external baselines.}
We verify the effectiveness of our semantic and spatial conditioning through an ablation study. First, we consider encoding part conditions naively. Rather than using our proposed Part Layout Adapter, we add the part labels to the global text prompt and encode the spatial conditions with a Fourier + linear embedding. Next, we consider using the adapter to encode all parts together rather than independently, providing all labels and bounding boxes to the adapter at once. Finally, we remove spatial conditions from Stage 2 and include only semantic conditions during multi-part generation, as per CubePart~\cite{cubepart}. We present paired generation results to a vision-language model (VLM)~\cite{gpt51} and 27 study participants, instructing them to assess their geometric quality (\eg detail, lack of artifacts) and how well they align with the specified part semantics. Additionally, we assess part-level spatial alignment by asking users how well the parts of the model align with a bounding box overlay, and we compute the Intersection over Union (IoU) between the bounding box of each output part and its corresponding input bounding box. Further details are provided in supplementary.

As seen in \cref{fig:full-ablation}, the naive and entangled encodings can result in overlapping parts and artifacts as the model struggles to distinguish between parts, and removing spatial conditions from Stage 2 can result in missing parts as the model must rely only on semantics to perform part segmentation. In contrast, our full method greatly improves part alignment and slightly improves geometric quality in most cases, as demonstrated by the evaluation (\cref{fig:full-quantitative}). Qualitative results are presented in \cref{fig:full-qualitative,fig:additional-results}.

\changes{The ablation above isolates our design choices. We now compare our full pipeline against external baselines: FullPart~\cite{fullpart} and OmniPart~\cite{omnipart}. We supply a ground-truth rendering alongside the ground-truth boxes, while our method receives only the text prompt and layout. Conditioning and matching otherwise follow Sec.~\ref{sec:stage-2-evaluation}. Results on PartObjaverse-Tiny are given in Tab.~\ref{tab:endtoend}. 
MultiCube matches FullPart~\cite{fullpart} on layout compliance while
winning on both semantic alignment and geometric quality. The comparison
favours FullPart on Box IoU by construction: it generates each part
independently and then rescales it to fit its assigned box, so high compliance is enforced by an explicit placement step rather than learned. MultiCube generates all parts jointly with no such step, yet reaches the same level of compliance. FullPart~\cite{fullpart} also takes no part labels, leaving part identities uncontrolled, which is reflected in its semantic alignment scores and in Fig.~\ref{fig:stage2-qualitative}. OmniPart attains a much lower Box IoU here. In Sec.~\ref{sec:stage-2-evaluation} it is initialized from ground-truth voxels, while here without that initialization its generated parts deviate substantially from its own conditioning, consistent with the difficulty it has conforming to input masks. 
}

\begin{table}[t]
  \caption{\textbf{End-to-end evaluation on PartObjaverse-Tiny.} Box IoU
  measures compliance with the input part layout. The VLM columns report MultiCube's win
  rate against each baseline on part semantic alignment and geometric quality,
  with ties counted as half wins. Both baselines
  additionally receive a ground-truth rendering; OmniPart also receives a
  ground-truth 2D segmentation. MultiCube is conditioned on text and layout
  alone.}
  \label{tab:endtoend}
  \centering
  \small
  \setlength{\tabcolsep}{1.8pt}
    \begin{tabular}{lcccc}
    \toprule
    & & \multicolumn{2}{c}{\small MultiCube win rate} \\
    \cmidrule(lr){3-4}
    End-to-end & Box IoU $\uparrow$ & VLM Sem. & VLM Qual. \\
    \midrule
    GT render + FullPart      & \textbf{0.723} & 82.9\%  & 89.0\%  \\
    GT render/seg + OmniPart  & 0.343 & 84.9\%  & 82.2\%  \\
    \midrule
    MultiCube (end-to-end)    & \underline{0.717} & --- & --- \\
    \bottomrule
    \end{tabular}
\end{table}

\input{figTex/additional-results}

\subsection{Applications}

\paragraph{Fully Automated Pipeline with LLM-generated Layouts.} To reduce manual effort, users can prompt large language models (LLMs) to generate bounding box layouts for an input prompt instead of specifying the layout manually (\cref{fig:teaser} (c), \cref{fig:full-qualitative}). Although the generated layouts are sometimes imperfect, our method is robust to small bounding box inaccuracies and gracefully handles imperfections; for example, it produces a correct shape despite incorrectly oriented wheels on the stroller input in \cref{fig:full-qualitative}, top-left.

\input{figTex/scene-generation}

\paragraph{Scene Generation.} Although our method is trained primarily on object-level shapes, it generalizes to full scenes, as shown in \cref{fig:scene-generation}.

\paragraph{Part-Level Customization.} 
By generating structurally disentangled meshes with explicit part-level conditions, MultiCube naturally supports generation-based editing workflows. Users can perform targeted part-level customizations by modifying the input schema or layout, such as adding or removing a bounding box, altering a specific part's text prompt, or adjusting its spatial dimensions. Although each operation triggers a full re-generation of the object, our strict per-part conditioning ensures that the overall structural layout and the semantic identities of the unmodified components are reliably preserved, as demonstrated in \cref{fig:teaser} (b).

\paragraph{Articulation.} 
Beyond static editing, the structural decomposition achieved by MultiCube facilitates downstream animation. Traditional monolithic 3D generation requires tedious manual post-processing to enable any form of movement. In contrast, because our method explicitly disentangles objects into semantically labeled components with clearly defined spatial bounding boxes, the generated distinct meshes are immediately ready for rigid-body articulation. As demonstrated in  \cref{fig:teaser} (b), users can easily assign kinematic properties and pivot points based on the semantic roles of individual parts, enabling realistic animations.

%% file: figTex/full-ablation.tex
\begin{figure}[tb]
    \centering
    \includegraphics[width=\linewidth]{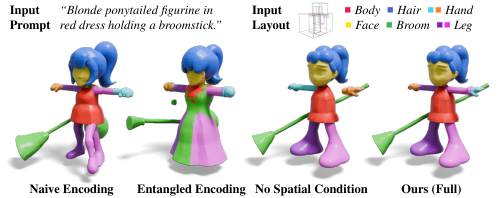}
    \caption{\textbf{Qualitative ablation results.} We compare our full model with the following ablated setups: a naive part condition encoding with no adapter, an entangled part condition encoding, and using semantic conditions only during multi-part generation.}
    \label{fig:full-ablation}
\end{figure}

%% file: figTex/full-quantitative.tex
\definecolor{wincolor}{HTML}{3cb44b} %
\definecolor{tiecolor}{HTML}{cccccc} %
\definecolor{losscolor}{HTML}{e6194b} %

\pgfplotsset{
    ablation axis style/.style={
        xbar stacked,
        xmin=0, xmax=100,
        xtick={25, 50, 75},
        xticklabels={25\%, 50\%, 75\%},
        ytick={1,2,3},
        font=\scriptsize,
        width=6.7cm, 
        height=3.7cm,
        enlarge y limits=0.3,
        title style={align=center, text width=6.5cm, yshift=-2mm}, %
        bar width=12pt,
        clip=false,
        nodes near coords={\pgfmathprintnumber{\pgfplotspointmeta}},
        nodes near coords align=center,
        every node near coord/.append style={
            font=\scriptsize, %
            text=white,
            /pgf/number format/fixed,
            /pgf/number format/precision=0
        },
    }
}

\newif\ifshowlegend

\newcommand{\ablationchart}[3]{
    \pgfplotstableread[row sep=crcr]{
        y win tie loss\\
        #3
    }\datatable

    \begin{tikzpicture}
        \begin{axis}[
            ablation axis style,
            title={\textbf{#1}},
            #2
        ]
        
        \addplot[fill=wincolor, draw=none, point meta=explicit, every node near coord/.append style={text=black}] 
            table[x expr={\thisrow{win}/(\thisrow{win}+\thisrow{tie}+\thisrow{loss})*100}, y=y, meta=win] {\datatable};
        \ifshowlegend \addlegendentry{Win} \fi
        
        \addplot[fill=tiecolor, draw=none, point meta=explicit, every node near coord/.append style={text=black}] 
            table[x expr={\thisrow{tie}/(\thisrow{win}+\thisrow{tie}+\thisrow{loss})*100}, y=y, meta=tie] {\datatable};
        \ifshowlegend \addlegendentry{Tie} \fi
        
        \addplot[fill=losscolor, point meta=explicit, draw=none] 
            table[x expr={\thisrow{loss}/(\thisrow{win}+\thisrow{tie}+\thisrow{loss})*100}, y=y, meta=loss] {\datatable};
        \ifshowlegend \addlegendentry{Loss} \fi
        
        \end{axis}
    \end{tikzpicture}
}

\begin{figure*}[!tbp]
    \centering
    \begin{subfigure}{0.33\linewidth}
        \centering
        \showlegendtrue
        \ablationchart{Part Semantic Alignment (VLM Eval)}
        {
            yticklabels={No Spatial\\Condition, Entangled\\Encoding, Naive\\Encoding},
            yticklabel style={text width=1cm, align=right}, %
            legend to name=sharedlegend,
            legend columns=-1,
            legend style={
                font=\scriptsize, %
                draw=none, 
                /tikz/every even column/.append style={column sep=0.5cm}
            },
            xticklabels={,,}
        }
        { %
            1 119 136 44 \\
            2 112 119 68 \\
            3 121 105 73 \\
        }
    \end{subfigure}%
    \hfill%
    \begin{subfigure}{0.33\linewidth}
        \centering
        \ablationchart{Part Spatial Alignment (Box IoU)}
        {
            yticklabels={,,},
            xticklabels={,,}
        }
        { %
            1 225 0 74 \\
            2 274 0 25 \\
            3  129 0 170 \\
        }
    \end{subfigure}%
    \hspace{-1.5mm}%
    \begin{subfigure}{0.33\linewidth}
        \centering
        \ablationchart{Geometric Quality (VLM Eval)}
        {
            yticklabels={,,},
            xticklabels={,,}
        }
        { %
            1 139	73	87\\
            2 155	8	136\\
            3 156	12	131\\
        }
    \end{subfigure}

    \vspace{-5mm}

    \begin{subfigure}{0.33\linewidth}
        \centering
        \showlegendfalse
        \ablationchart{Part Semantic Alignment (Human Eval)}
        {
            yticklabels={No Spatial\\Condition, Entangled\\Encoding, Naive\\Encoding},
            yticklabel style={text width=1cm, align=right}, %
        }
        { %
            1 18	15	4\\
            2 11	22	3\\
            3 7	26	5\\
        }
    \end{subfigure}%
    \hfill%
    \begin{subfigure}{0.33\linewidth}
        \centering
        \showlegendfalse
        \ablationchart{Part Spatial Alignment (Human Eval)}
        {
            yticklabels={,,},
        }
        { %
            1 18 16 3\\
            2 15 16 4 \\
            3 6 28 4\\
        }
    \end{subfigure}%
    \hspace{-1.5mm}%
    \begin{subfigure}{0.33\linewidth}
        \centering
        \showlegendfalse
        \ablationchart{Geometric Quality (Human Eval)}
        {
            yticklabels={,,},
        }
        { %
            1 11 23 3\\
            2 11 8  17\\
            3 17 11 10\\
        }
    \end{subfigure}

    \vspace{-3mm}
    
    \ref*{sharedlegend}
    
    \vspace{-1mm}
    
    \caption{\textbf{Ablation study.} We run head-to-head comparisons of our model against ablated setups on part-level alignment and geometric quality.}

    \label{fig:full-quantitative}
\end{figure*}

%% file: figTex/stage2-qualitative.tex
\begin{figure*}[!tbp]
    \centering
    \includegraphics[width=\linewidth]{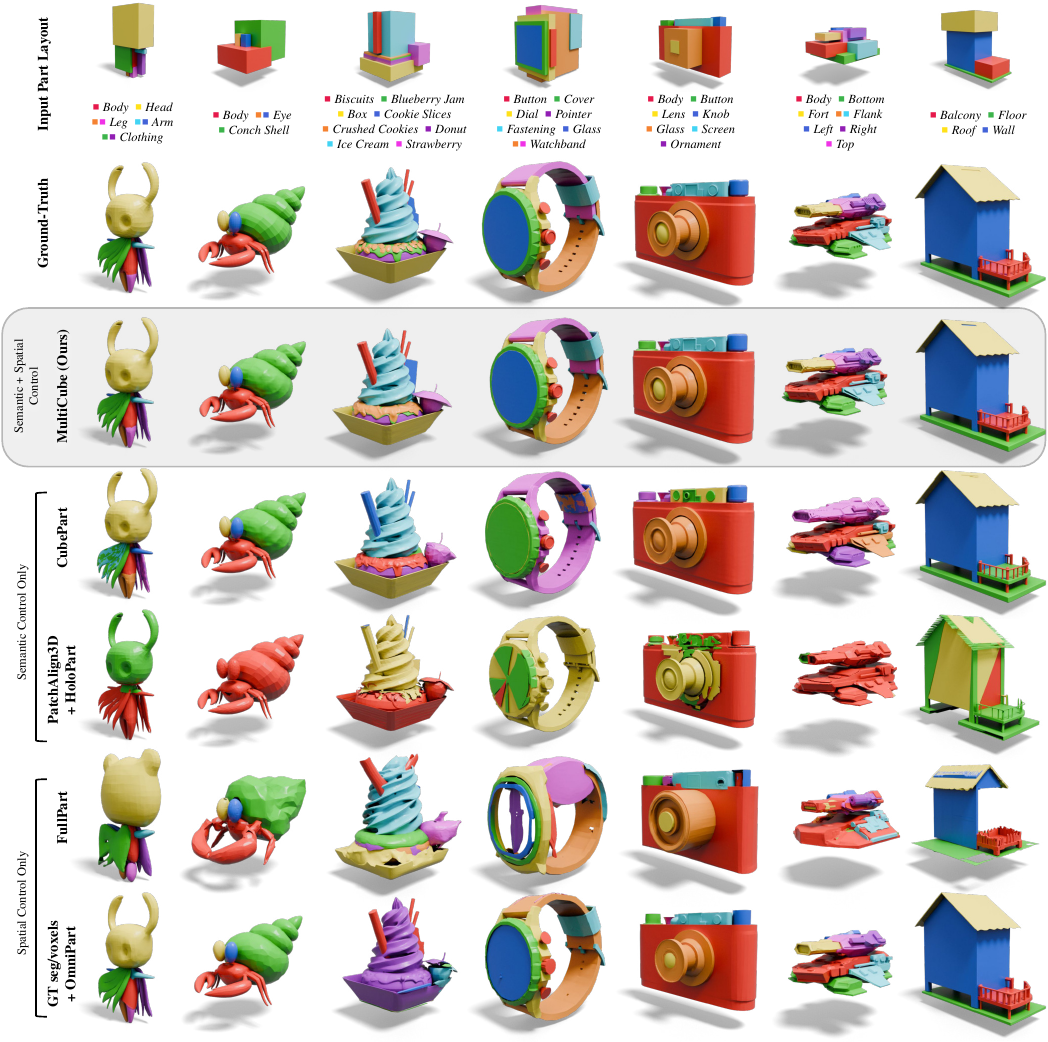}
    \caption{\textbf{Qualitative multi-part generation comparison.} Our method conditions on an input shape, spatial layout, and semantic part labels. CubePart~\cite{cubepart} and the ``PatchAlign3D~\cite{patchalign3d} + HoloPart~\cite{holopart}'' pipelines condition on shape + semantic part labels. FullPart~\cite{fullpart} and OmniPart~\cite{omnipart} are image- and spatial layout-conditioned; OmniPart additionally requires 2D segmentation and 3D voxel inputs. Compared with baselines, our method exhibits cleaner part decompositions and more detailed geometry.}
    \label{fig:stage2-qualitative}
\end{figure*}

%% file: tables/stage2-quantitative.tex
\begin{table}[tb]
    \caption{\textbf{Quantitative evaluation for multi-part generation.} We compute Chamfer distance and F-scores at part and holistic levels, alongside bounding box alignment at the part level. Our method strongly improves part-level accuracy. Note that OmniPart is given a \textit{ground-truth} 2D segmentation and 3D voxels as input.}
    \label{tab:stage2-quantitative}
    \centering
    \small
    \setlength{\tabcolsep}{1.2pt}
    \begin{tabular}{@{}lccccc@{}}
        \toprule
        \multirow{2}{*}{PartObjaverse-Tiny} & \multicolumn{3}{c}{Part-Level} & \multicolumn{2}{c}{Holistic-Level} \\
        \cmidrule(lr{0.1mm}){2-4} \cmidrule(lr{0.1mm}){5-6}
         & CD $\downarrow$ & F-score $\uparrow$ & Box IoU $\uparrow$ & CD $\downarrow$ & F-score $\uparrow$ \\
        \midrule
        PartCrafter & 0.612 & 0.346 & 0.176 & 0.142 & 0.781  \\
        PartPacker & 0.325 & 0.594 & 0.321 & 0.084 & 0.925 \\
        \midrule
        GT seg/vox + OmniPart & \underline{0.058} & \underline{0.940} & \underline{0.753} & \textbf{0.038} & \textbf{0.991} \\
        FullPart & 0.118 & 0.828 & 0.723 & 0.105 & 0.871 \\
        \midrule
        PA3D + HoloPart & 0.228 & 0.698 & 0.405 & 0.048 & 0.971 \\
        CubePart & 0.183 & 0.787 & 0.580 & \underline{0.044} & \underline{0.978} \\
        \midrule
        MultiCube (Ours) & \textbf{0.040} & \textbf{0.948} & \textbf{0.847}  & 0.046 & 0.967 \\
        \bottomrule
    \end{tabular}
\end{table}

%% file: figTex/full-qualitative.tex
\begin{figure*}[t]
    \centering
    \includegraphics[width=\linewidth]{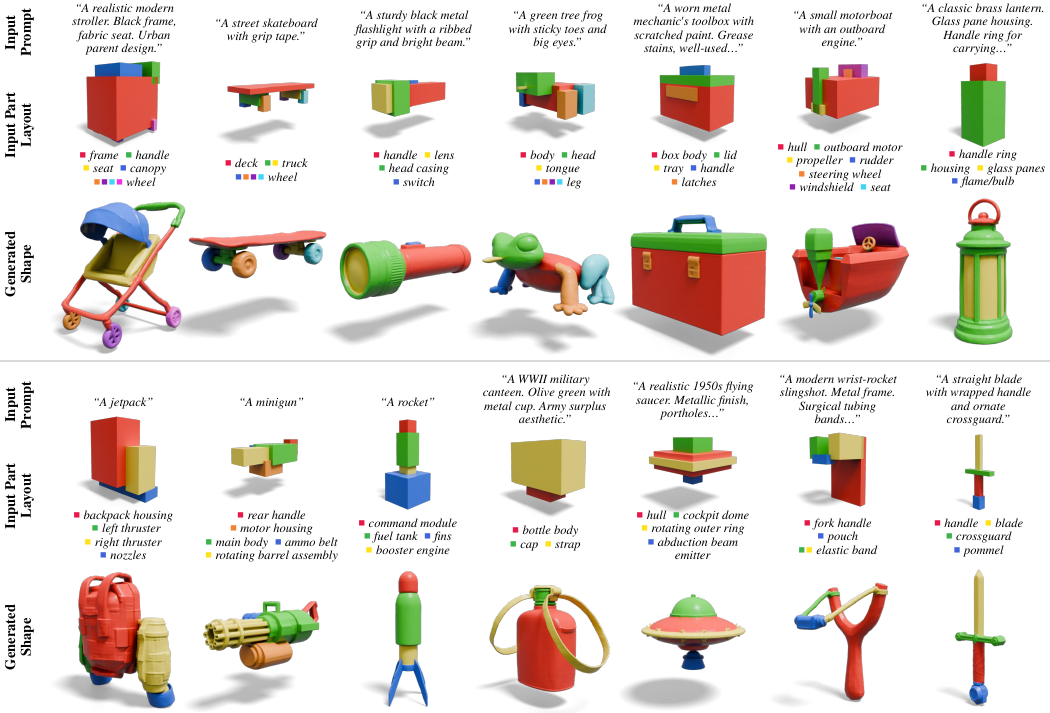}
    \caption{\textbf{Full pipeline generation results.} MultiCube generates compositional 3D objects from input text prompts and part layouts. Here, the part bounding box layouts are \textit{automatically generated} using GPT-5.1~\cite{gpt51}; hence, the generation process from prompt to output shape is fully automatic.}
    \label{fig:full-qualitative}
\end{figure*}

%% file: figTex/additional-results.tex
\begin{figure*}[t]
    \centering
    \includegraphics[width=\linewidth]{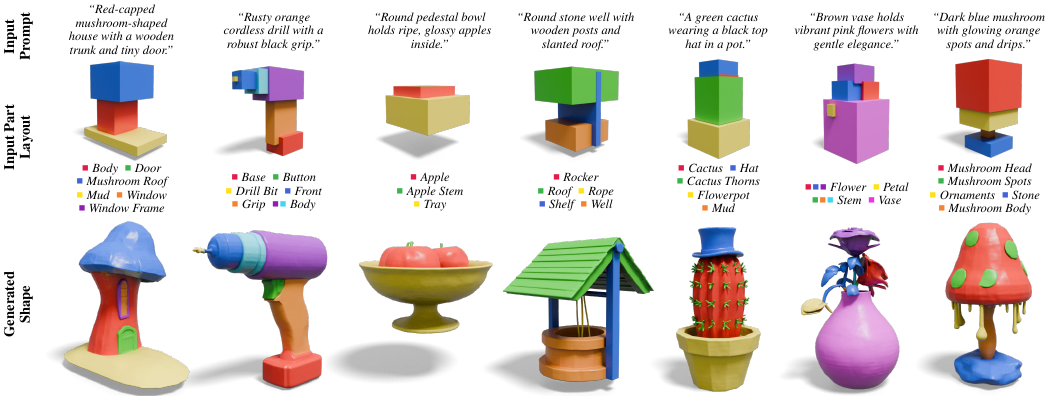}
    \caption{Additional generation results, using part layouts from PartObjaverse-Tiny.}
    \label{fig:additional-results}
\end{figure*}

%% file: figTex/scene-generation.tex
\begin{figure}[tb]
    \centering
    \includegraphics[width=\linewidth]{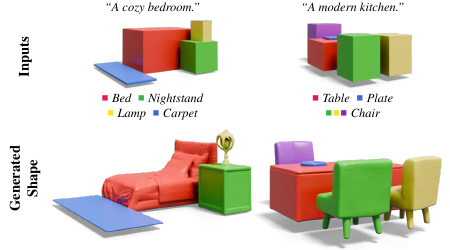}
    \caption{\textbf{Generalization to scenes.} Though our model is trained mostly on object-level shapes, it generalizes to producing layout-conditioned scenes. Users can control the position, size, and type of each object in the scene.}
    \label{fig:scene-generation}
\end{figure}

%% file: sections/9_conclusion.tex
\section{Discussion and Limitations}

\input{figTex/failure-cases}

We have presented MultiCube, a novel method for generating compositional 3D meshes with part-level semantic and spatial control. By utilizing a two-stage approach with a dedicated Part Layout Adapter that effectively encodes part-level conditions, we achieve high-quality, controllable multi-part mesh generation that surpasses baselines. While MultiCube is a strong step toward fully controllable 3D asset creation, several avenues for improvement remain:

\textbf{Failure cases.} MultiCube is robust to small bounding box perturbations; however, if the bounding boxes are severely mis-specified, the generated shape may be strange (\cref{fig:failure-cases}, left). Additionally, while Stage 2 utilizes cross-part attention and spatial conditioning to maintain global consistency, the model occasionally generates colliding parts (\cref{fig:failure-cases}, right). Future work could integrate collision-avoidance loss terms to ensure parts remain spatially distinct.

\textbf{Iterative editing.} Currently, MultiCube generates the full output at once; it does not support ``generate-and-tweak'' loops where individual parts are modified without regenerating the entire assembly. Extending our framework to support local, iterative editing would increase its utility in production environments.

\section*{Acknowledgments}

We thank Gaurav Parmar, Maxwell Jones, Nupur Kumari, Ruihan Gao, Sheng-Yu Wang, Michael Palleschi, Maurice Chu and Kiran Bhat for their feedback and helpful discussions. Ava Pun is supported by the Packard Fellowship, the CGRS D funded by the Natural Sciences and Engineering Research Council of Canada (NSERC), and the IITP grant funded by the Korean Government (MSIT) (No. RS-2024-00457882, National AI Research Lab Project). This work was partly done while Ava Pun was interning at Roblox.

%% file: figTex/failure-cases.tex
\begin{figure}[tb]
    \centering
    \includegraphics[width=0.9\linewidth]{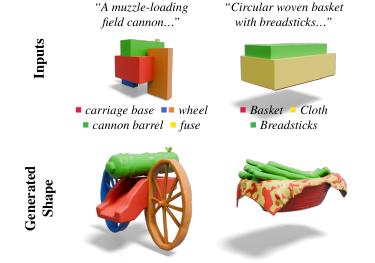}
    \caption{\textbf{Failure cases.} Our model may produce strange outputs if the input bounding boxes are wildly incorrect (left). Additionally, it occasionally produces overlapping parts (right).}
    \label{fig:failure-cases}
\end{figure}

%% file: sections/A_supplemental.tex
\appendix

\onecolumn

\section*{Appendix}

\section{Additional Evaluation Details}

We provide additional details on our evaluation method (\cref{sec:experiments}).

\paragraph{Custom Dataset Generation}
The full LLM text prompt used to generate bounding box layouts for our custom evaluation dataset is given in \cref{lst:bboxes-prompt}.

\input{figTex/mv-render-user-study}

\paragraph{VLM Evaluation}
The VLM prompts used to evaluate part semantic alignment and geometric quality are given in \cref{lst:semantic-prompt} and \cref{lst:quality-prompt}, respectively. For each prompt, the VLM receives four multi-view renders of each of the two 3D models to be evaluated (\cref{fig:mv-render}), and performs a head-to-head comparison. The parts in the models are distinguished by different solid colors. The order of the given models is randomized.

\paragraph{Human Evaluation}
We present human evaluators with two interactive 3D models side-by-side, alongside a text prompt and colored part labels (\cref{fig:user-study}). The models are shown in random order. We ask evaluators the following questions to assess part semantic alignment, part spatial alignment, and geometric quality, respectively:

\begin{itemize}
    \item \textbf{Which model better captures the given part labels?}\\
    Each part should clearly represent the part label with the same color.\\
    $\square$ Model A \ $\square$ Model B \ $\square$ Tie
    \item \textbf{Which model better follows the specified bounding box layout?}\\
    Each part should fit neatly within the bounding box of the same color.\\
    $\square$ Model A \ $\square$ Model B \ $\square$ Tie
    \item \textbf{Which model has higher-quality 3D geometry?}\\
    More details, fewer artifacts, fewer intersecting parts, etc.\\
    $\square$ Model A \ $\square$ Model B \ $\square$ Tie
\end{itemize}

\changes{
Twenty-seven participants took part in the study, drawn from researchers in graphics, vision, and machine learning. The study comprised 136 randomly ordered comparison pairs. Participants rated as many pairs as they chose, with a mean of 5.4 pairs each. Each pair received between one and three ratings. Model order was randomized within each pair, and no participant was shown which method produced which model.
}

\paragraph{Baseline Input Details}
The baseline methods used during Stage 2 evaluation (\cref{sec:stage-2-evaluation}) accept varying kinds of inputs: part labels, monolithic shapes, monolithic voxels on a $64^3$ grid, part bounding boxes, 2D images, and 2D segmentations. These inputs are specified in \cref{tab:baseline-inputs}. Notably, OmniPart is given a \textit{ground-truth} part segmentation as input, enhancing its ability to decompose a given 3D object into parts.

\lstset{basicstyle=\footnotesize\ttfamily, breaklines=true, breakatwhitespace=true, literate={`}{\textasciigrave}1}
\begin{lstlisting}[float=*,caption={Prompt for generating part bounding box layouts.},label={lst:bboxes-prompt}]
You are an expert 3D technical artist and spatial reasoning engine. Your task is to generate 3D bounding boxes for specific parts of an object based on a text prompt.

### 1. Coordinate System & Scale
Assume a normalized 3D space where the entire object fits within a bounding volume of [-1.0, 1.0] across all axes. 
- **Origin (0, 0, 0):** The center of the entire object.
- **X-axis (Width):** Left (-1.0) to Right (+1.0).
- **Y-axis (Height):** Bottom (-1.0) to Top (+1.0). Ground level is at Y = -1.0.
- **Z-axis (Depth):** Back (-1.0) to Front (+1.0).

### 2. Output Format
For each part, you must provide a bounding box defined by two 3-tuples (represented as arrays of 3 floats):
1.  **"dimensions"**: [size_x, size_y, size_z] representing the width, height, and depth of the box.
2.  **"center"**: [x, y, z] representing the 3D coordinates of the exact center of the box.

### 3. Spatial Rules
- **Adjacency & Geometry:** Parts must connect logically based on the object's real-world structure. For example, a "head" should sit directly on top of a "body" (their Y-boundaries should touch). "Wheels" or "legs" should touch the ground plane (Y = -1.0). Bounding boxes may intersect, but there should be no floating or disconnected bounding boxes.
- **Symmetry:** If a part implies multiple symmetrical components (e.g., "wheels" or "wings") but is provided as a single label, provide a single bounding box that encompasses all of them (e.g., a wide box at the bottom for "wheels" spanning the X-axis).
- **Proportions:** Ensure the dimensions make sense relative to one another (e.g., a torso is generally larger than a head).

### 4. Input
- **Prompt:** "{prompt}"
- **Parts:** {parts}

### 5. Task
First, write a brief <thinking> section where you spatially map out the object. Determine the relative sizes and vertical/horizontal stacking of the parts to ensure they logically connect and don't awkwardly intersect.
Then, output a pure JSON object enclosed in ```json blocks with the following structure:

```json
{
  "parts": {
    "part name 1": {
      "dimensions": [w, h, d],
      "center": [x, y, z]
    },
    "part name 2": {
      "dimensions": [w, h, d],
      "center": [x, y, z]
    }
  }
}
```
\end{lstlisting}

\begin{lstlisting}[float=*,caption={Prompt for evaluating part semantic alignment.},label={lst:semantic-prompt}]
You are an expert 3D model evaluator. You are presented with two images containing multi-view renders of two 3D models. The first image contains Model A, and the second contains Model B. The models are generated from the text prompt "{prompt}", and they are meant to include the following parts: {part 1}, {part 2}, .... These parts correspond to the colors {color 1}, {color 2}, ..., respectively.

**Task:** Determine which model better adheres to the given part labels.

**Criteria:**
1. Does the {color 1} part clearly represent the part label: "{part 1}"?
2. Does the {color 2} part clearly represent the part label: "{part 2}"?
...

**Output:** Provide a brief step-by-step analysis of how well each model adheres to the part labels. Then, state your final verdict by explicitly outputting a JSON file containing a single key "verdict" with the value being either "Model A", "Model B", or "Tie". For example:
```json
{
  "verdict": "Model A"
}
```
\end{lstlisting}

\begin{lstlisting}[float=*,caption={Prompt for evaluating shape quality.},label={lst:quality-prompt}]
You are a quality assurance tester for 3D assets. You are presented with two images containing multi-view renders of two 3D models. The first image contains Model A, and the second contains Model B. Each model contains several parts, indicated by different solid colors. You are evaluating the two models based on their geometric quality.

**Task:** Determine which model is of higher 3D quality.

**Criteria:**
1. **Parts:** Are the parts distinct from each other with clean boundaries, or do they intersect/overlap?
2. **Detail:** Does the model have clear and distinct details, or does it look overly smooth and blurry?
3. **Artifacts:** Are there messy deformations or floating/disconnected geometry?

**Output:** Provide a brief step-by-step analysis of the visual quality of both models. Then, state your final verdict by explicitly outputting a JSON file containing a single key "verdict" with the value being either "Model A", "Model B", or "Tie". For example:
```json
{
  "verdict": "Model A"
}
```
\end{lstlisting}

\input{tables/baseline-inputs}

%% file: figTex/mv-render-user-study.tex
\begin{figure*}[b]
\begin{minipage}{0.49\linewidth}
    \centering
    \begin{subfigure}{0.48\linewidth}
        \centering
        \includegraphics[width=\linewidth]{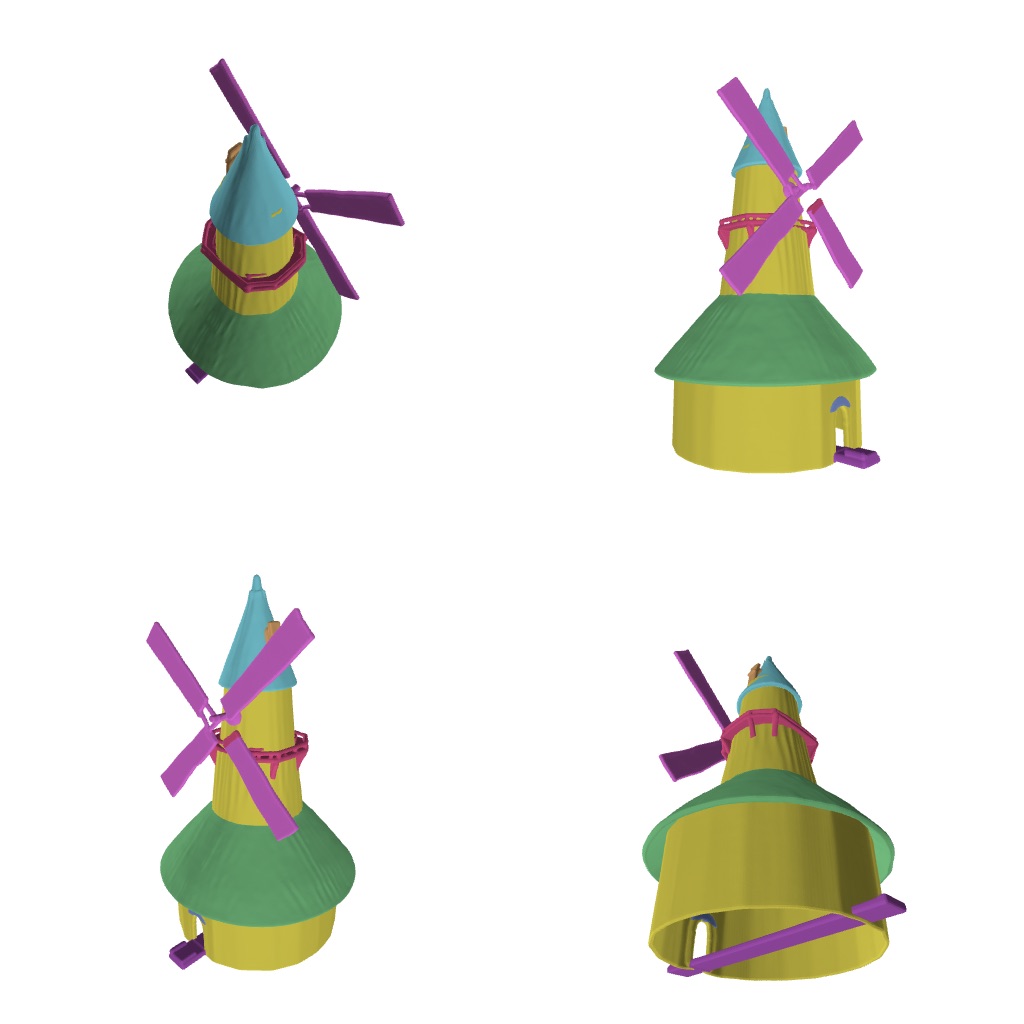}
        \caption{Model A}
    \end{subfigure}%
    \hfill%
    \begin{subfigure}{0.48\linewidth}
        \centering
        \includegraphics[width=\linewidth]{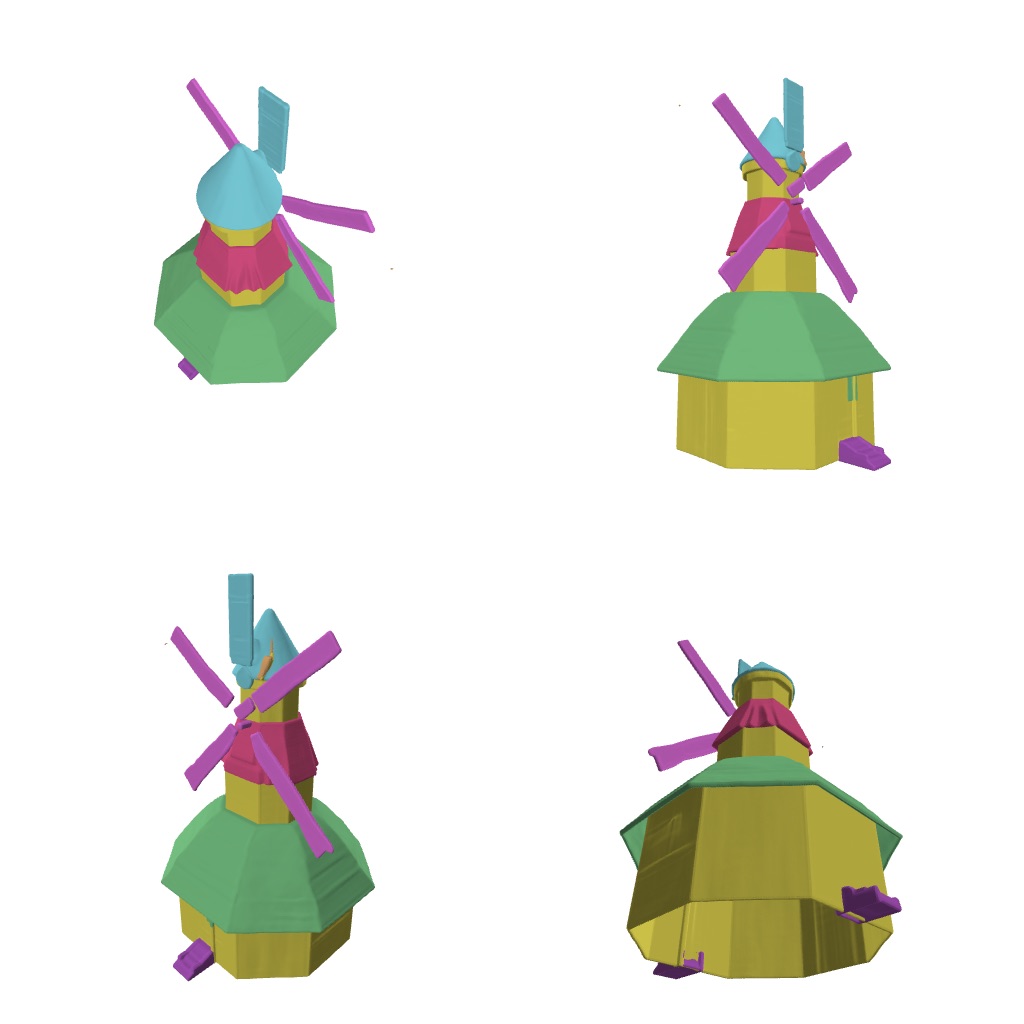}
        \caption{Model B}
    \end{subfigure}
    \caption{We present a VLM with multi-view renders of each of two models, asking it to determine which has better part semantic alignment and geometric quality.}
    \label{fig:mv-render}
\end{minipage}%
\hfill%
\begin{minipage}{0.49\linewidth}
    \centering
    \includegraphics[width=\linewidth]{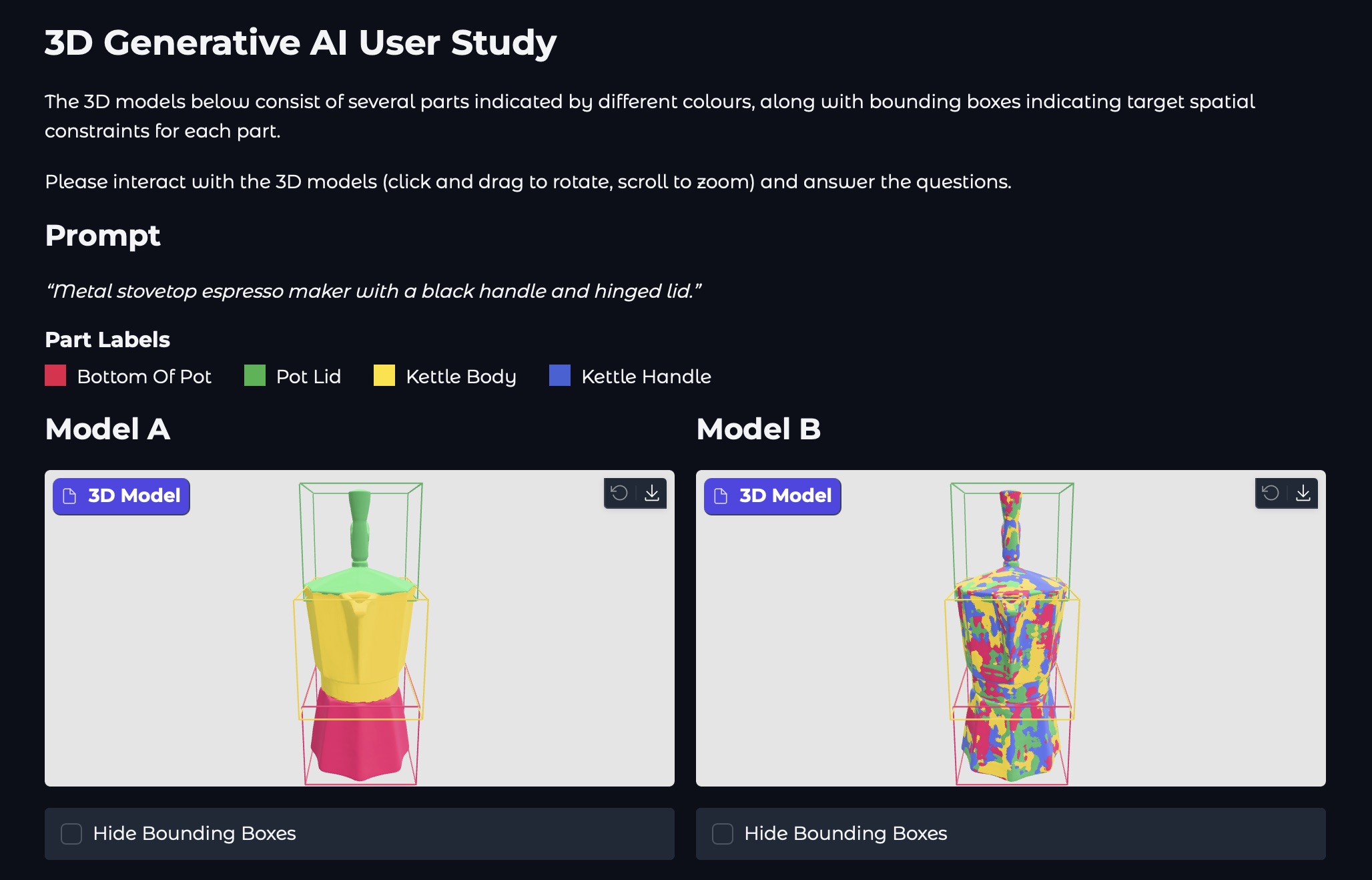}
    \caption{The interface used for head-to-head comparisons in the human user study.}
    \label{fig:user-study}
\end{minipage}
\end{figure*}

%% file: tables/baseline-inputs.tex
\begin{table*}[!tbp]
    \caption{\textbf{Inputs to multi-part generation methods.} Inputs include ground-truth part labels, shapes, voxelized shapes, part bounding boxes, rendered 2D images, and 2D segmentations from PartObjaverse-Tiny.}
    \label{tab:baseline-inputs}
    \centering
    \begin{tabular}{@{}lcccccc@{}}
        \toprule
        \multirow{3}{*}{Method} & \multirow{2}{*}{\makecell{Input Part\\Labels}} & \multicolumn{3}{c}{3D Inputs} & \multicolumn{2}{c}{2D Inputs}\\ 
        \cmidrule(lr{0.1mm}){3-5} \cmidrule(lr{0.1mm}){6-7}
         && \makecell{Monolithic\\Shape} & \makecell{Monolithic\\Voxels} & \makecell{Part \\Boxes} & \makecell{Rendered\\Image} & \makecell{Part\\Seg} \\
        & {\scriptsize \makecell{\textit{Grip, Body,}\\\textit{Magazines,}\\\textit{Barrel, Tail}}} & \includegraphics[width=1.4cm,valign=c]{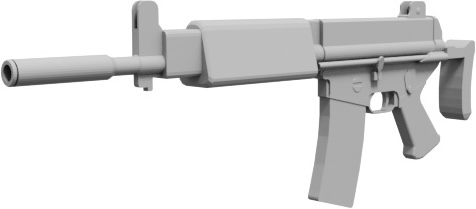} & \includegraphics[width=1.4cm,valign=c]{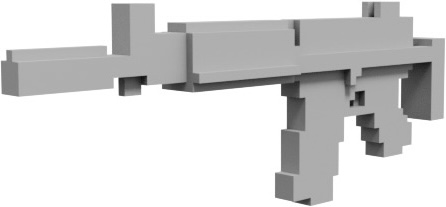} &
        \includegraphics[width=1.4cm,valign=c]{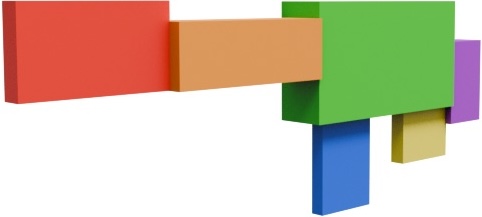} &
        \includegraphics[width=1.4cm,valign=c]{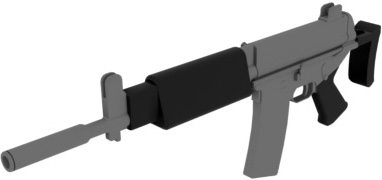} &
        \includegraphics[width=1.4cm,valign=c]{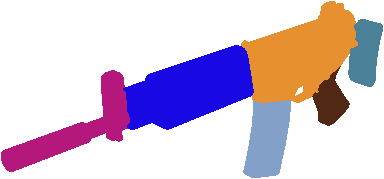} \\
        \midrule
        PartCrafter & &&&& \checkmark & \\
        PartPacker & &&&& \checkmark & \\
        \midrule
        GT seg/voxels + OmniPart & && \checkmark & \checkmark & \checkmark & \checkmark \\
        FullPart & && & \checkmark & \checkmark & \\
        \midrule
        PatchAlign3D + HoloPart & \checkmark & \checkmark & & & &  \\
        CubePart & \checkmark & \checkmark & & & &  \\
        \midrule
        MultiCube (Ours) & \checkmark & \checkmark & & \checkmark & &  \\
        \bottomrule
    \end{tabular}
\end{table*}